\documentclass[12pt]{article}
\usepackage{amsmath,amssymb}
\usepackage{array}
\usepackage{stmaryrd}
\usepackage{cancel}
\usepackage{braket}
\usepackage{comment}
\usepackage{graphicx}
\usepackage{caption}
\usepackage{color}
\usepackage{dcolumn}
\usepackage{bm}
\usepackage[numbers,comma,sort&compress]{natbib}
\usepackage[leftcaption]{sidecap} 
\sidecaptionvpos{figure}{c}

\newcolumntype{P}{>{\centering\arraybackslash}m{4.6cm}}
\newcolumntype{M}{>{\centering\arraybackslash}m{6cm}}
\newcolumntype{L}{>{\centering\arraybackslash}m{5.5cm}}

\title{Magic state distillation and gate compilation in quantum algorithms for quantum chemistry}
\date{January 6, 2015}

\author{Colin J. Trout$^{1}$ and Kenneth R. Brown$^{1,2,3}$\\
$^{1}$\emph{\footnotesize School of Chemistry and Biochemistry, Georgia Institute of Technology, Atlanta, GA 30332}\\
$^{2}$\emph{\footnotesize School of Physics, Georgia Institute of Technology,  Atlanta, GA 30332}\\
$^{3}$\emph{\footnotesize School of Computational Science and Engineering, Georgia Institute of Technology, Atlanta, GA 30332}}

\begin{document}

\maketitle

\begin{abstract}
Quantum algorithms for quantum chemistry map the dynamics of electrons in a molecule to the dynamics of a coupled spin system. In order to reach chemical accuracy for interesting molecules, a large number of quantum gates must be applied which implies the need for quantum error correction and fault-tolerant quantum computation.  Arbitrary fault-tolerant operations can be constructed from a small, universal set of fault-tolerant operations by gate compilation. Quantum chemistry algorithms are compiled by decomposing the dynamics of the coupled spin-system using a Trotter formula, synthesizing the decomposed dynamics using Clifford operations and single-qubit rotations, and finally approximating the single-qubit rotations by a sequence of fault-tolerant single-qubit gates. Certain fault-tolerant gates rely on the preparation of specific single-qubit states referred to as magic states. As a result, gate compilation and magic state distillation are critical for solving quantum chemistry problems on a quantum computer. We review recent progress that has improved the efficiency of gate compilation and magic state distillation by orders of magnitude.
\end{abstract}

\clearpage

\section*{\sffamily \Large INTRODUCTION}

Quantum simulation is the mapping of one quantum system to a second, well-controlled quantum system in order to both understand how the system changes as parameters are varied and to allow measurements that may be hard to access in the original system \cite{SimPhysComp,UnivQuantSim, QuantSimRev}.  Quantum simulation can offer an advantage over classical computer simulation, since there is no compact way to represent arbitrary quantum states with classical bits. The application of quantum simulation to quantum chemistry is seen as a promising application for a future quantum computer. 

Calculations have shown that a full-configuration interaction (FCI) calculation on a quantum computer can be solved with a number of operations that is polynomial in the number of spin-orbitals assuming that an approximation to the quantum state of interest can be prepared \cite{SimQCEnergy}. The bulk of the algorithm consists of simulating the dynamics of the second-quantized molecular Hamiltonian with $N$ spin-orbitals:
\begin{equation} \label{eq:MolHamilt}
\hat{\mathcal{H}}=\sum_{j,k} h_{jk}\hat{c}^\dagger_j \hat{c}_k+\sum_{j,k,l,m}  h_{jklm}\hat{c}^\dagger_j\hat{c}^\dagger_k\hat{c}_l\hat{c}_m
\end{equation}
where $h_{jk}$ are the single-electron energies corresponding to the kinetic energy and the attraction to the nuclei, $h_{jklm}$ are the two-electron energies due to electron-electron interaction, and $\hat{c}^\dagger_j (\hat{c}_j)$ is an electron creation (annihilation) operator for the spin-orbital $j$.
The dynamics are simulated for a set of  times $t,\, 2t,\, 4t,... \, 2^kt=T$ where $t$ sets the maximum energy that can be observed and $T$ sets the precision due to the standard limits of the Fourier transform \cite{UnivQuantSim,BrownPRL2006}. 

\begin{SCfigure}
\includegraphics[width=1.75in,keepaspectratio=true]{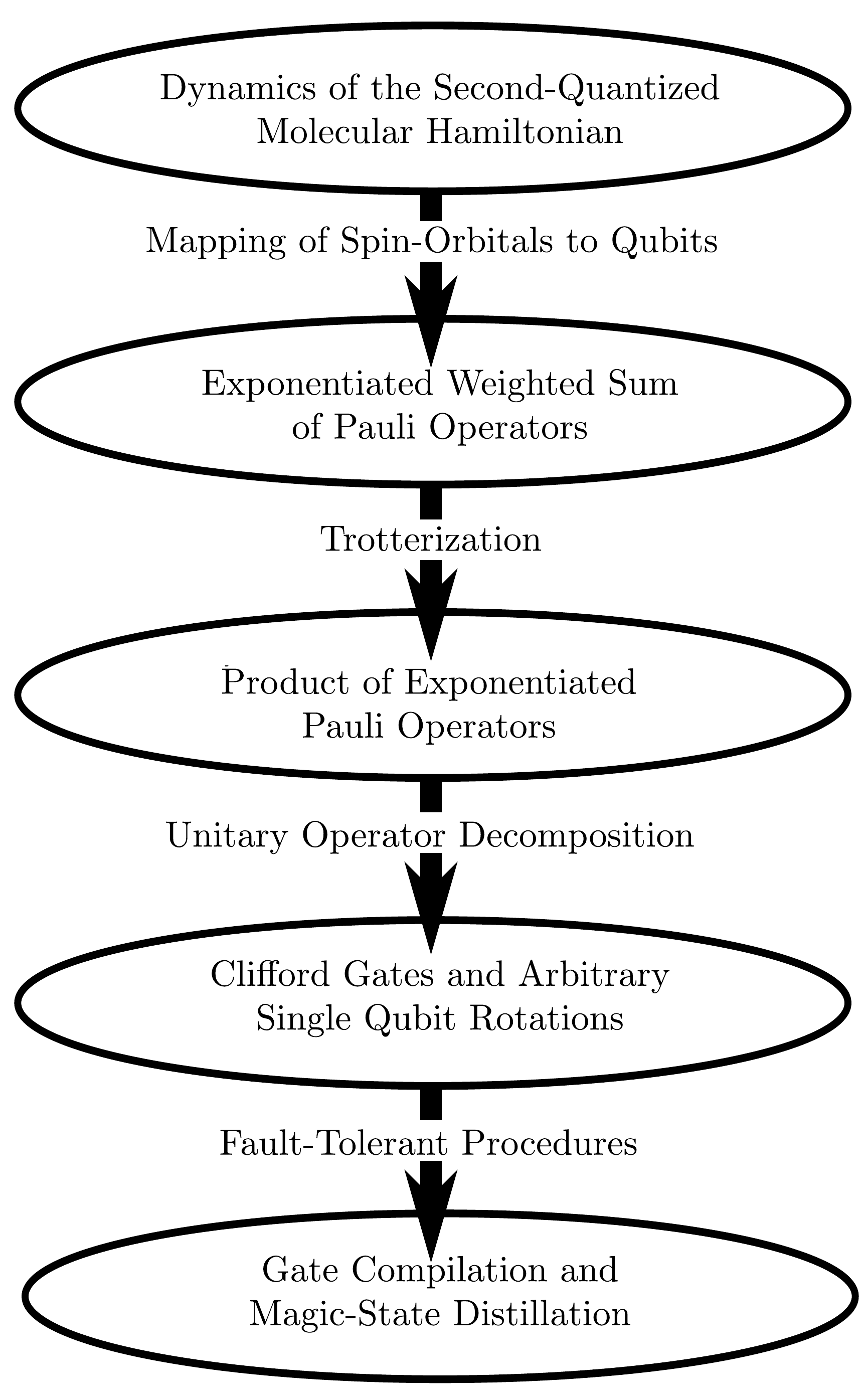}
\caption{\label{fig:compilationflow} Presented are the steps to simulate the dynamics of a Hamiltonian on a quantum computer. This review focuses on recent progress in the final step consisting of gate compilation and magic-state distillation.}
\end{SCfigure}

The quantum simulation is performed by mapping the dynamics of the molecular Hamiltonian to a series of quantum gates (Figure \ref{fig:compilationflow}).  The first step is to map fermionic operators to spin operators \cite{SimFermiUnivQC}. There are two generic methods of performing this mapping: Jordan-Wigner, which encodes orbital occupation in individual spins at the cost of $N$-body spin operators to account for the exchange symmetry  \cite{JordanWigner, JordanWignerQC, SimQCEnergy}, and Bravyi-Kitaev, which uses $\log(N)$-body spin operators at the cost of losing the direct mapping from orbitals to qubits \cite{BravyiKitaev2002, BKtransformQuantChem}.  For restricted Hamiltonians, e.g., models where only spatially local interactions between electrons are allowed, simpler mappings are possible \cite{SimFermiUnivQC}.

A universal quantum computer can generate an arbitrarily accurate approximation to any unitary evolution on 2$^n$ two-level systems or qubits.  A key result of quantum computation is that this unitary operation can be broken into arbitrary single-qubit gates and any entangling two-qubit gate \cite{BarencoPRA1995, DiVincenzoPRA1995}. In this framework, the compilation of various unitary operations can remain challenging.

For quantum chemistry, the simulation of each $M$-body spin operator is broken down into operations on one and two qubits, which can be combined using the Trotter formula, $\exp\left(-\frac{it}{\hbar}\sum_k{\hat{\mathcal{H}}_k}\right) \approx \left(\prod_k \exp\left(-\frac{it}{\hbar}\hat{\mathcal{H}}_k/m\right)\right)^m$.  Based on estimates of the required Trotter time-steps and the choice of spin-transformation, the total number of operations scales as $N^8$ \cite{GateCountSmallQuantComp}, for arbitrary $h$ values, to $N^{6}$ using Jordan-Wigner \cite{TrotterStepAccurateQuantSim} or $N^5\log(N)$ using Bravyi-Kitaev \cite{ExploitLocalQuantSimQuantChem} for values of $h$ commonly seen in molecules.  This is quite promising for FCI since the scaling within the regime of worst-case coupled-cluster singles and doubles (CCSD), $N^6$, \cite{CCSD} and the number of required quantum bits scales only as $N$. We also note that there is a connection between electron correlation and the required number of Trotter-steps resulting in quantum algorithms with only $N^4$ total operations for molecules well-described by the Hartree-Fock approximation \cite{TrotterStepAccurateQuantSim}.

These estimates assume perfect qubits and operations. Actual qubits have errors requiring fault-tolerant methods for robust quantum computation. Fault-tolerant quantum error correction encodes a single logical qubit into multiple physical qubits. This allows us to preserve information in the logical level by detecting and correcting errors at the physical level  \cite{KeyIdeasQECC}. Operations on logical qubits must be fault-tolerantly performed, which requires the careful design of logical gates from physical gates. Fortunately, only a restricted set of gates is required to achieve universal quantum computation \cite{MikeNIke}. For many error-correcting codes, the fault-tolerant gates fall into two classes: gates that can be directly applied to the logical qubit and gates that require the consumption of a specially prepared ancillary state, known as a magic state \cite{UnivQuantCompCliffordnNoisyAncillas}. 

The operators that were decomposed assuming no quantum error correction now need to be decomposed into this restricted gate set. The effect of this decomposition on the total number of qubits and gates has been considered previously for simulations of spin and molecular Hamiltonians \cite{TIMSteane,FastQuantSimFTQC,TIMSurface}, but in the last few years there has been a remarkable reduction in the required resources. In this review, we will  examine advances in the compilation of arbitrary single-qubit unitary gates from restricted gate sets and improvements in the generation of magic states by a process called distillation, where many low-quality magic states are converted into a few high-quality magic states in a fault-tolerant manner. 

\section*{\sffamily \Large Quantum Gates and Teleportation}

We start with some useful notation and definitions. For further details, we recommend the textbook Ref. \cite{MikeNIke}.  The Pauli matrices on a single-qubit $j$ are represented by $\hat{X}_j=\hat{\sigma}^x_j$, $\hat{Y}_j=\hat{\sigma}^y_j$, and $\hat{Z}_j=\hat{\sigma}^z_j$. The two common bases are the computational or $z$ basis corresponding to eigenstates of $\hat{Z}$, $\hat{Z}\ket{0}=\ket{0}$ and $\hat{Z}\ket{1}=-\ket{1}$, and the $x$ basis corresponding to eigenstates of $\hat{X}$, $\hat{X}\ket{+}=\ket{+}$ and $\hat{X}\ket{-}=-\ket{-}$.  The Hadamard matrix is a self-adjoint matrix that transforms between these bases, $\hat{H}\hat{X}\hat{H}=\hat{Z}$.  The standard two-qubit gate is the $\hat{\mathrm{CNOT}}(i,j)$ which adds the value of qubit $i$ in the computational basis to qubit $j$ modulo 2 and can be written as 
\begin{equation}
\hat{\mathrm{CNOT}}(i,j)\equiv\ket{0}\bra{0}_i\hat{I}_j +\ket{1}\bra{1}_i\hat{X}_j
\end{equation}
where $\hat{I}_j$ is the identity on the $j$th qubit.

Two important groups of unitary operations are the Pauli group and the Clifford group. The Pauli group, $\mathcal{P}$, is generated by multiplication of individual Pauli matrices on each qubit.  The Clifford group, $\mathcal{C}$, is the group of unitary transformations that maps the Pauli group to itself; for $\hat{g} \in \mathcal{C}$, $\hat{g}\mathcal{P}\hat{g}^\dagger = \mathcal{P}$. On a single-qubit, the Clifford transformations correspond to the symmetry elements of the chiral octahedron (the point group $O$).  The Clifford group can be generated by the single-qubit operators $\hat{H}$, $\hat{S}=\ket{0}\bra{0}+i\ket{1}\bra{1}$, and the two-qubit operator $\hat{\mathrm{CNOT}}$.

Each element of the second-quantized molecular Hamiltonian can be mapped to a sum of elements in $\mathcal{P}$ \cite{SimElecStruct}.  The dynamics can be generated using only the Clifford group which encode the creation/annihilation operators of the molecular Hamiltonian and arbitrary single-qubit rotations which encode the electron energy $\left(h_{jk}\right)$ dependent time-evolution.  For an explicit example of such a decomposition, see the  Appendix.  The coefficients in the molecular Hamiltonian can vary greatly from molecule to molecule and, at first, it seems that arbitrary rotations are required.  We can remove this requirement by noting that arbitrary rotations can be generated from the repeated application of only two rotations. Almost any pair of rotations will work with the exception of two rotations that are symmetry elements of the same point group. The canonical choice in quantum information is $\hat{H}$ and $\hat{T}=\exp(i\pi/8)\hat{R}_Z(\pi/4)$ also referred to as the $\pi/8$ gate. The $\hat{T}$-gate, which does not preserve the chiral octahedron, plus the Clifford gates is a common universal gate set. Note that $\hat{T}^4=\hat{S}^2=\hat{Z}$. It has long been known that any rotation can be efficiently simulated using these gates \cite{ClassQuantComp} but, as we discuss in this review, recent work has drastically reduced the cost.

\begin{figure}
\begin{center}
\includegraphics[width=0.9\columnwidth,keepaspectratio=true]{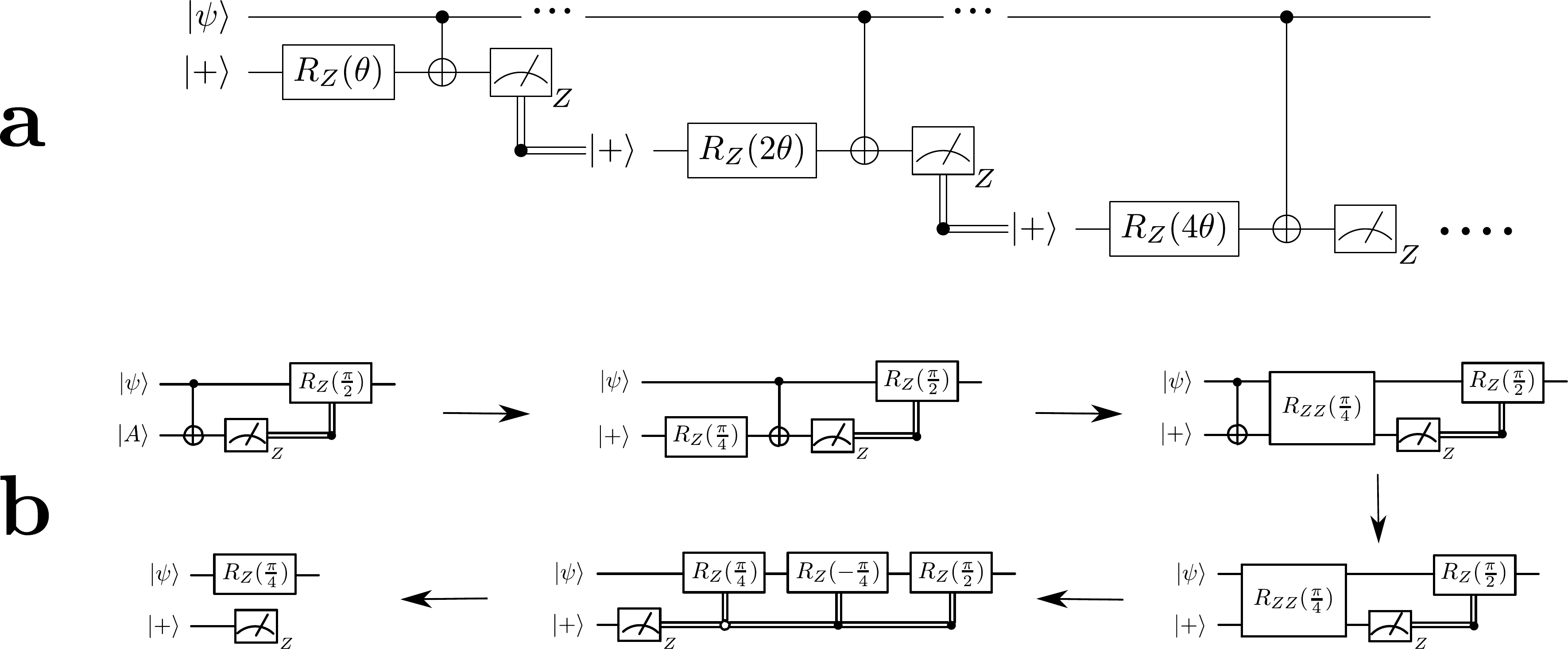}
\end{center}
\caption{\label{fig:circuitidentities} In this figure, we use the language of quantum circuits \cite{MikeNIke} to show the implementation of general rotations, $\hat{R}_{Z}(\theta)$, and the deterministic implementation of the rotation $\hat{R}_{Z}(\pi/4)$ by the use of gate teleportation. Single black horizontal lines correspond to qubits.  Horizontal double black lines to classical bits. Vertical lines represent control with a black (white) dot signifying control on the qubit or bit being 0 (1). Time flows from left to right.  (a)  Circuit for the implementation of the gate $\hat{R}_{Z}(\theta)$.  Teleportation is attempted by the application of the rotation $\hat{R}_{Z}(\theta)$ to the ancillary state $\ket{+}$, a $\hat{\mathrm{CNOT}}$ to induce entanglement between the input state $\ket{\psi}$ and the ancillary state $\ket{+}$, and the measurement of the ancillary state in the computational basis.  If the measurement outcome is $0$, then the desired gate was applied and the circuit terminates as indicated by the perforations of the quantum wire containing the input state.  If the outcome of the measurement is instead $1$, then the rotation was applied in the undesired direction about the axis and a correction rotation, $\hat{R}_{Z}(2\theta)$ , may be attempted in a similar manner. We may continue to apply these corrections until we get the desired measurement result.  (b) Deterministic teleportation of the rotation, $\hat{R}_{Z}(\pi/4)$, to the input register $\ket{\psi}$ from the utilization of magic state, $\ket{A}=\ket{A(\pi/4)}$.  The first circuit consists of the unknown data state $\ket{\psi}$, the magic state $\ket{A}=\ket{A(\pi/4)}$, a $\hat{\mathrm{CNOT}}$, a measurement in the $z$ basis, and a measurement dependent Clifford operation, $\hat{R}_{Z}(\pi/2)$. In the second circuit we write $\ket{A}$ in terms of the $\ket{+}$ and the gate $\hat{R}_{Z}(\pi/4)$. In the third circuit, we have pushed the single-qubit gate through the $\hat{\mathrm{CNOT}}$, transforming it into a two-qubit unitary $\hat{R}_{ZZ}(\pi/4)$. In the fourth circuit, the $\hat{\mathrm{CNOT}}$ can be removed since the identity and the NOT operator, $\hat{X}$, have the same action on $\ket{+}$. In the fifth circuit, we note that the $Z$ measurement commutes with the operator $\hat{R}_{ZZ}$ and transforms $\hat{R}_{ZZ}$ into a measurement controlled $\hat{R}_{Z}$. Finally, we observe that the measurement value does not change the overall operator applied to the first qubit.}
\end{figure} 

Imagine a situation where qubit 1 can only have Clifford gates act on it (including the $\hat{\mathrm{CNOT}}$) and qubit 2 can have Clifford gates and rotations around the $z$-axis by an arbitrary angle $\hat{R}_Z(\theta)$. We can effectively perform $\hat{R}_Z (\theta)$ on the first qubit by using the important quantum primitive of gate teleportation \cite{TeleportGottesman1999}. The procedure, depicted in figure \ref{fig:circuitidentities}a, works as follows:  qubit 1 is in an arbitrary state. Qubit 2 is prepared in the state $\ket{+}=1/\sqrt{2}\left(\ket{0}+\ket{1}\right)$ and the rotation $\hat{R}_Z(\theta)$ is applied. We will refer to this state as  $\ket{A(\theta)}$. Next $\hat{\mathrm{CNOT}}(1,2)$ is applied and then qubit 2 is measured in the computational basis. The net effect is as follows: if qubit two is measured in state $\ket{0}$, then $\hat{R}_Z\left(\theta\right)$ was applied to qubit 1. If qubit two is measured in state $\ket{1}$, then $\hat{R}_Z\left(-\theta\right)$ was applied to qubit 2. In this case, we can gamble and attempt to apply $\hat{R}_Z(2\theta)$ and continue this procedure until success \cite{RUSSingleQubit}. 

If $\theta=\pi/4$, we do not have to gamble. If we measure qubit 2 to be in the state 1, we have performed the undesired rotation $\hat{R}_z(-\pi/4)$. This can be corrected deterministically by applying the Clifford group operator $\hat{R}_Z(\pi/2)=e^{-i\pi/4}\hat{S}$ to the first qubit (see figure \ref{fig:circuitidentities}b). We can then use our ability to perform $\hat{R}_Z(\pi/4)$ to create a deterministic circuit for implementing $\hat{R}_Z(\pi/8)$. This process can continue and it is often useful to consider gates that are of the form $\hat{R}_Z(\pi/2^k)$ as well.

For a physical qubit, for example a trapped ion or the nuclear spin of a molecule, the difference between rotations of different angles corresponds to simply a change in the pulse area.  There is no greater difficulty in applying $\hat{R}_Z(\pi/32)$ or $\hat{Z}$.  Many quantum systems have shown single-qubit gates with error probabilities on the order of 10$^{-4}$ or less per gate \cite{NMRSingleQubit,PenningSingleQubit, OpticalLatticeOlmschenk,KentonSingleQubit,SuperconductorThesholdBarends2014, SmallCrosstalkPiltz, IonLowErrorGates2014}.  If these error rates can also be achieved with two-qubit gates, then a quantum algorithm with  $\gtrsim 10^{4}$ gates will not complete successfully, limiting the application of quantum computers to quantum chemistry. The solution to this problem is fault-tolerant quantum error correction.

\section*{\sffamily \Large Quantum Error Correction and Preferred Universal Gate Sets}

For classical computers, error correction is possible due to the discrete nature of the states and the ability to copy information. At first glance, quantum computers and error correction seem mismatched since quantum states form a continuous space and the copy of quantum systems is forbidden by the no-cloning theorem \cite{NoCloning}.  The insight that allows for quantum error correction is that, instead of having discrete states, one can have discrete errors \cite{KeyIdeasQECC}.  Each error shifts the data from one subspace of the system to another.  A set of measurements can be applied that reveals which subspace is occupied without disturbing the data. The collapse of the state onto specific subspaces after measurement transforms the possible range of continuous errors into a set of discrete errors.

For adequate partitioning into subspaces that can distinguish errors, a single logical qubit is encoded into multiple physical qubits.  If the probability of an error on each physical gate is below a certain threshold, then arbitrarily long computations can be performed if sufficiently powerful error correction is used \cite{ThresholdAharonov1997, ThresholdKnill1998, ThresholdAliferis2006}.  The threshold theorem is a key result of quantum information and the value of the threshold depends strongly on the codes and error models but, for realistic noises and qubit geometries, it has been calculated to be as high as $\approx10^{-2}$ \cite{SurfaceCodeRaussendorf2007,SurfaceCodeTutorialFowler2012}.  Currently two physical systems, trapped-ion qubits and superconductor qubits have demonstrated a set of universal operations and measurements with error rates below this threshold \cite{SuperconductorThesholdBarends2014, IonLowTwoQubitGates2014}.

When considering a code, both the resources required for memory and for computation are pertinent. For every code, there is a small set of gates that can be implemented transversally. Transversal operations act on each physical qubit separately in the logical qubit.  This implies that the resource overhead of transversal operations is equivalent to the cost of the encoded memory in terms of the numbers of physical qubits.  A common example is where the logical operation is the same physical operation on all qubits. Some examples of quantum error-correcting codes and their transversal gate sets can be found in \cite{TransversalExamples}. For many codes these correspond to the Clifford operations, for some codes only the Pauli operations, and for others only rotations of certain angles about the Z axis. For single-qubits, ignoring the global phase, these operations correspond respectively to the point groups $O$, $D_2$, and $D_{2^k}$. 

For an important family of codes the Clifford gates are transversal and the costly gate is the $\hat{T}$ gate \cite{FowlerCC}.  This gate can be implemented by teleportation if the magic state $\ket{A(\pi/4)}=\hat{T}\ket{+}$ can be prepared following figure \ref{fig:circuitidentities}b.  At first glance this seems like an avoidance of the problem.  Clearly to generate $\ket{A(\pi/4)}$, we need to apply $\hat{T}$.  However, we can use a different code family to build a distillation circuit that allows us to efficiently generate high-quality $\ket{A(\pi/4)}$ states from multiple noisy copies.

\subsection*{\sffamily \large $\hat{{\bf T}}$ Gate Distillation}

Bravyi and Kitaev developed the first method for $\hat{T}$ gate distillation based on the  $\llbracket 15,1,3 \rrbracket$ quantum Reed-Muller code to distill the magic state $\ket{A(\pi/4)}$ \cite{UnivQuantCompCliffordnNoisyAncillas}.   This code takes 15 noisy ancillary magic states, with error rate $p$, and encodes them into 1 qubit of information with a failure in the distillation circuit when 3 or more errors occur during the encoding.  By performing the error check measurements and corrections in the quantum error correcting routine, one may iteratively project the noisy set of magic states onto one higher fidelity magic state, with an output error rate $\mathcal{O} \left(p^3\right)$, resulting in a noisy to distilled state ratio of 15-to-1 \cite{UnivQuantCompCliffordnNoisyAncillas}.  

\begin{figure}
\begin{center}
\includegraphics[width=0.7\columnwidth,keepaspectratio=true]{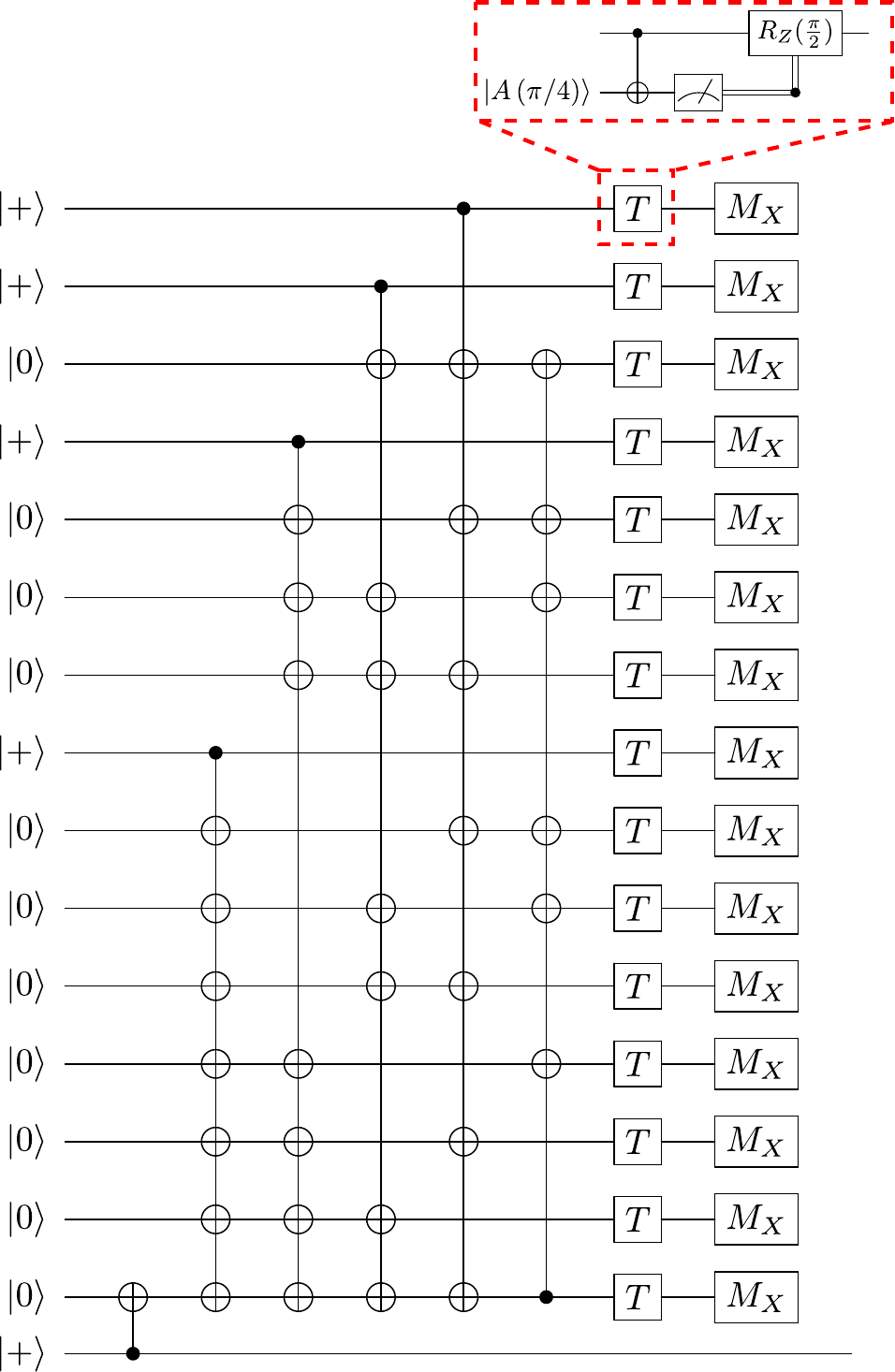}
\end{center}
\caption{\label{fig:BK15to1} The Bravyi-Kitaev 15-to-1 magic state distillation protocol.  The input to the circuit is 16 easy to prepare input states, $\ket{0}$ and $\ket{+}$, and 15 low-quality magic states with error $p$, and the output is 1 magic state with an output error rate of $\mathcal{O}\left(p^3 \right)$.  The circuit is composed of Clifford operations and Pauli measurements. It implements  15 deterministic $\hat{T} $ gate teleportations (figure \ref{fig:circuitidentities}b).  Subsequent rounds of distillation can be performed on the output states to suppress the errors of the magic state to arbitrary precision, thus allowing fault-tolerant implementation of the non-Clifford $\hat{T}$ gate.} 
\end{figure}

 The distillation circuit for this protocol is depicted in figure \ref{fig:BK15to1}. Notice the circuit employs only Clifford operations except for the input magic states used for deterministic $\hat{T}$ gate teleportation as in Fig. \ref{fig:circuitidentities}b.  Further rounds of distillation can be performed on distilled states to suppress error rates up to an arbitrary precision and, thus, allows for fault-tolerant implementation of the non-Clifford $\hat{T}$ gate. 

Meier, Eastin, and Knill (MEK) utilized a  $\llbracket 10,2,2 \rrbracket$ error detection routine within the distillation circuit \cite{MagicState4Qubit}.  This code has a higher ratio of noisy to distilled states of 10-to-2 at the cost of a reduction in error suppression of the distilled states of $\mathcal{O} \left(p^2 \right)$. Dependent on the target error rate for the distilled state, the 10-to-2 technique can more efficiently produce distilled states than the 15-to-1 protocol.  There is a regime of low input error rates for which the quadratic suppression of errors from 10-to-2 technique is sufficient and extra resource cost for cubic error suppression via the Bravyi-Kitaev distillation is deemed unnecessary.  For high-initial errors and low target errors,  initial implementation of the 15-to-1 protocol for maximum error suppression followed by the 10-to-2 protocol to minimize ancilla overhead can result in a favorable resource scaling relative to using a single protocol \cite{MagicState4Qubit}.

Bravyi and Haah introduced a method with a ratio of noisy to distilled states of $3k+8$ to $k$ for $\ket{A(\pi/4)}$ with output error rates comparable to that of Meier-Eastin-Knill, $\mathcal{O} \left(p^2 \right)$ \cite{MagicStateLowOverhead}.  Furthermore, they introduced a systematic method of generating distillation protocols by realizing an equivalence between a set of quantum error correcting codes that admit a transversal $\hat{T}$ gate and a family of matrices known as triorthogonal matrices.  The nonzero elements of these matrices in a row denote the set of simultaneous checks performed on the physical qubits per step of the error correction routine.  Therefore, generating new distillation routines translates into generating new triorthogonal matrices.  As with the 10-to-2 distillation, the utility of this technique is substantial at lower error rates and, therefore, admits serial use of distillation protocols due to constraints on the input error regime in which the resource gains of this technique outweighs the gain in output state precision  \cite{MagicStateLowOverhead}.

\section*{\sffamily \large Gate Decomposition Techniques and Minimizing Non-Clifford Rotations}

We will now present three general methods for generating fault-tolerant arbitrary unitary transformations.  The first method uses a minimal, universal, elementary gate set (typically Clifford operations and $\hat{T}$ gates) and approximates arbitrary rotations through successive application of gates from this set. These methods minimize the number of $\hat{T}$ gates to reduce the overhead due to distillation. The second method, known as complex distillation, incorporates an overcomplete gate set which is composed of Clifford operations and incremental rotations about the $x$, $y$, or $z$ axis on the Bloch sphere to construct arbitrary rotations. The third method is sequential probabilistic application of gates through teleportation.   Table \ref{tbl1} illustrates the decrease is resource costs from the progress in these techniques.


\subsubsection*{Optimization with a Minimal Gate Set}

Gate decomposition techniques critically rely on a fundamental result in quantum computing known as the Solovay-Kitaev theorem. Given a set of single-qubit gates (e.g., $\hat{H}$, $\hat{T}$, and $\hat{T^\dagger}$) that generate a ``dense" set of unitary gates, it is possible to approximate any unitary gate to an arbitrary precision $\left(\delta \right)$, by an efficient number of operations, $\propto \log^c(1/\delta)$ where $c$ is a constant \cite{ClassQuantComp}. The challenge is how to compile these elementary gates into accurate approximations of arbitrary unitary operations. 

We will now outline the progress in methods of approximating arbitrary unitary rotations from a set of Clifford and $\hat{T}$ gates.  The Solovay-Kitaev algorithm presents a step-wise procedure for generating arbitrary gates \cite{SK2005}.  First, a library is generated of all rotations up to a sequence length of elementary gates $l$. The target rotation is compared to the closest rotation in the library.  The difference between these rotations is a small rotation that can be generated from the library taking advantage of the non-commutivity of rotations. This can be done in a recursive manner to generate arbitrarily accurate gates without ancillary qubits.   Further optimization methods have been implemented to the Solovay-Kitaev algorithm with improvements that reduce the level of recursion required for the calculation of the sequences \cite{OptimSKAlgo}. The final elementary gate count also depends critically on the size of the library as demonstrated by Bocharov and Svore \cite{SKSequenceLengthBocharov2012}.

The best gate counts will occur for infinitely large libraries. However, this approach becomes impractical for high-precision approximations. For moderate precision, Fowler found optimal sequences by a ``brute force" space search that utilizes information about previously applied gate sequences to selectively screen for the next gate sequence \cite{OptimSKAlgo,ConstrSteaneCodeSingleLogQubit}.  The precision at which this method can approximate gates is limited as the space search is exponential in gate sequence length, but this result was nevertheless incredibly insightful as it illuminated the significant gap in gate sequence length between optimal decompositions and the Solovay-Kitaev result.  

An alternate approach for developing an optimal decomposition required the use of ancillary states that could offer small corrections to the approximated state through a method known as phase kickback \cite{AsymOptimalConstAncilla}. From these methods, a representation developed where the problem is mapped to a unit field and optimal gate sequences that can be obtained without the use of ancillary states \cite{AsymOptimalConstAncilla,EffCliffnTApprox,PracticalApproxUnitaries,OptimalCliffordTDecompOfZ}.  Non-deterministic methods of approximating such rotations have also been developed which make use of a minimal set of gates and ancillary states which are measured to iteratively project the input state to the required state within a given precision \cite{FastQuantSimFTQC,FloatingPoint,EffSynthRUS, PQFterminates}.

Significant progress has recently been sparked by work performed by Kluichnikov, Maslov, and Mosca (KMM) \cite{FastEfficientSynthesis}.  This work presented a rigorous proof that any single-qubit unitary operator can be \emph{exactly} decomposed into a set of single-qubit Clifford and $\hat{T}$-gates, with one $\ket{0}$ ancilla state, if and only if the matrix elements of the unitary operator belong to a special algebraic group and fulfill constraints on the assignment of the matrix elements to guarantee unitarity.  Furthermore, it was shown that this decomposition was efficient.  Giles and Selinger extended this proof to include $n$-qubit unitary transformations and eliminated the need for ancillary qubits in the exact decompositions \cite{ExactSynthMulti}.  This result is powerful as it added intuition into the search space of potential circuits resulting in a deviation from random circuit search methods.

The question remained: given a desired rotation within a given precision, how does one find a rotation with an exact decomposition that lies within that precision?  Furthermore, can one find an exact unitary decomposition, within that precision, which is minimal in the use of the costly $\hat{T}$ gate?  Various algorithms for searching the space of unitaries with exact decompositions have been implemented for the generation of these circuits \cite{AsymOptimalConstAncilla,EffCliffnTApprox,PracticalApproxUnitaries,OptimalCliffordTDecompOfZ}.  These methods differ from the search methods implemented by Fowler in that, instead of searching over gate sequences, they search for the entries in the single-qubit unitary that simultaneously satisfy the norm equation to guarantee unitarity and are members of an algebraic ring, the cyclotomic integers, to ensure the existence of an exact decomposition.  Initial attempts at these optimizations resulted in decompositions that did \cite{AsymOptimalConstAncilla} and did not \cite{EffCliffnTApprox} require the use of a few ancillary $\ket{0}$ states in the circuit. Additional algorithms resulted in an optimal solution in the number of $\hat{T}$ gates with ancilla \cite{PracticalApproxUnitaries}, and Ross and Selinger recently showed optimal solutions in gate decompositions that are exact and require no ancillary qubits \cite{OptimalCliffordTDecompOfZ}. Ironically, the efficiency of the decomposition depends on the ability to factor numbers, a task at which quantum computers excel \cite{ShorsAlgo}.  Fortunately, probabilistic versions of this search algorithm can be implemented in polynomial classical runtime and can achieve the third-to-optimal decomposition sequence, with $\hat{T}$ gate counts comparable to the optimal solution \cite{OptimalCliffordTDecompOfZ}.

\subsubsection*{The Utility of an Expanded Gate Set from Complex State Distillation}

Complex distillation protocols \cite{DistillNonstabState,CISCLessMagic,OneStepComplexDistillation} utilize an overcomplete elementary gate set consisting of Clifford operations and a set of rotational states about a given axis.  These rotations about an axis are produced by a distillation protocol that takes magic states, such as those constructed from the routines above, as an input and generates a ``harder" rotation.  While employing subsequent distillation procedures on distilled states appears to lower efficiency,  the rationale behind these techniques is founded in the fact that the minimization of the number of ancilla required per magic state does not necessarily imply more cost-efficient computation as the above protocols are ignorant to the compiling cost of the computation; an issue addressed by these complex distillation techniques.  Indeed, there is gate number threshold above which generating a ``difficult" rotation directly may benefit over successive applications of a set of cheaper rotations when approximating an arbitrary rotation. As an example of how this may reduce a gate sequence length for an arbitrary desired rotation, if one has access to the rotations $\hat{R}_Z(\pi/2^k)$, one can approximate any arbitrary $\hat{R}_Z(\theta)$ by digital compilation.  This can be especially useful for algorithms that require on-the-fly compilation of unitary gates based on measurement of qubits \cite{DynamicSetComputing, OneStepComplexDistillation}.

These  protocols differ in the expanded gate set used and the distillation used to generate the expanded magic states.  What they do hold in common is that the expanded set of rotations about an axis are generated recursively, $\left(\hat{R}\left(\theta_1 \right) ~\rightarrow \hat{R}\left(\theta_2 \right) ~\rightarrow...~\rightarrow \hat{R}_Z\left(\theta_k\right) \right)$, with finer rotations requiring more distillation steps for all techniques and these rotations are applied to data via gate teleportation.  

Duclos-Cianci and Svore \cite{DistillNonstabState} use an elementary gate set that includes Clifford gates and $\hat{R}_Y$ rotations.  The resource states used in the teleportation were constructed from distillation of $\ket{H_i}$ magic states, $\ket{H_i}=\hat{R}_Y(2\theta_i)\ket{0}$ where $\theta_i$ is defined by $\cot \theta_i=\cot^{i+1} \pi/8 $. Note that $\ket{H_0} = \ket{+}$.  An additional routine utilizing a set of additional rotations $\left\{\ket{\psi_i^{0}},\ket{\psi_i^{1}},\ket{\psi_i^{2}} \right\}$ generated Clifford circuits with $\ket{H_i}$'s as an input creates a larger, denser set of rotations partitioning the $xz$ plane. The work presented a systematic method of generating dense states about a plane on the Bloch sphere. The rotations generated by the magic states are not uniformly partitioned and  although they can be combined to generate any angle, there is no natural decomposition.

Landahl and Cesare modified the previous approach to complex distillation techniques in two ways: first they addressed the non-uniform partitioning of the $yz$ plane by performing  $\hat{R}_Z\left(\pi/2^k\right)$ rotations and also implemented a distillation routine with a ``top down" recursion \cite{CISCLessMagic}.
In the ``top down" recursion method, additional $\hat{R}_Z \left(\pi/2^{k-1} \right)$ input states are needed if the $\hat{R}_Z \left(\pi/2^{k}\right)$ teleportation is faulty and even less $\hat{R}_Z \left(\pi/2^{k-2} \right)$ states are required as they are applied upon consecutive failures of the previous two teleportation attempts which probabilistically reduces the number of traditionally distilled states required for gate teleportation because the teleportation must fail $k-3$ times before a $\hat{T}\ket{+}$ magic state must be prepared \cite{DistillNonstabState,CISCLessMagic}. The technique used shortened Reed-Muller codes that require input magic states of increasing quality to achieve the smallest rotations. 

Duclos-Cianci and Poulin overcame the limits of using Reed-Muller codes by the implementation of a modified 10-to-2 distillation circuit \cite{MagicState4Qubit}  to generate magic states $\ket{Y_k}=\hat{Z}\hat{S}\hat{H}\ket{A(2\pi/2^k)}$ for $\hat{R}_Y\left(\pi/2^k\right)$ rotations. Conditional on the measurement, the distillatoin procedure transforms two noisy $\ket{Y_k}$ states into two quadratically improved $\ket{Y_k}$ states. The method requires Clifford operations and the application of a parity measurement, which consumes 16 magic states of the type $\ket{Y_3}$ and 1 magic state of the type $\ket{Y_{k-1}}$ \cite{OneStepComplexDistillation}. The similarity of the distillation for all $k$ allows the procedure to work with magic states of a fixed initial accuracy regardless of $k$.  For an input accuracy of 1\%, $\ket{0}$, a Clifford state, serves as a sufficiently accurate input state. The authors find a slightly lower resource overhead in terms of non-Clifford input states when substituting $\ket{0}$ for input $\ket{Y_k}$ when $k>8$ than for $k>3$. This method has comparable resource costs relative to $\hat{T}$-gate only compilation methods for arbitrary rotations, but substantial savings for rotations by angles $2\pi/2^k$.

\subsection*{\sffamily \large Non-Deterministic Application of Rotations}
The third set of methods can be thought as random-walk with correction teleportation techniques.  Teleportation is attempted with ancillary states that are measured to detect the success of the application of the gate, similar to the teleportation scheme shown in figure \ref{fig:circuitidentities}a.  If the gate fails, recursive corrections, such as those we have seen from complex distillation techniques, can be applied until the desired gate is applied.  An early example of this technique was shown by Jones et al. which implemented programmable ancilla rotations (PARS) \cite{FastQuantSimFTQC}.  This method required non-Clifford ancillary states that were originally expensive to generate but distillation techniques have made the production of Fourier states efficient \cite{DistillFourierStates}.  Recent methods have eliminated the need for these complex ancillary states to non-deterministically achieve arbitrary rotations \cite{EffSynthRUS,FloatingPoint}. The most recent non-deterministic technique by Bacharov et al. utilizes the knowledge of previous rotation attempts in a manner that probabilistically guarantees the termination of the circuit after a finite set of gates \cite{PQFterminates}.

\begin{table}[htbp]
\begin{tabular}[t]{|P|M|c|c| }
\hline
\textbf{Method} & $\hat{\mathrm{\textbf{T}}}$ {\bf Gate Count for Precision $\delta$} & \multicolumn{2}{c|}{\textbf{Non-Clifford States}}  \\ \hline
Solovay-Kitaev$^{a}$ \cite{SK2005} & $s_1 \left( \mathrm{log}^{3.97} \, \left(1/\delta\right)  \right)$ & $86$ & $8.0 \times 10^{5}$ \\ \hline
Fowler Search$^{a}$ \cite{SKSequenceLengthBocharov2012,ConstrSteaneCodeSingleLogQubit}  & $2.95 \, \mathrm{log}_{2} \left(1/\delta \right) +3.75 $ & $1.3 \times 10^2$ & $3.4 \times 10^{2 \, \dag}$ \\ \hline
PARs$^{b}$ \cite{FastQuantSimFTQC} & $s_2 \left( \mathrm{log}\left(1/\delta \right)\right)$ & $6.0 \times 10^2$ & $1.4 \times 10^{3 \, \dag \dag}$ \\ \hline
KMM$^{a}$ \cite{PracticalApproxUnitaries}  & $3.067 \, \mathrm{log}_{10} \left(1/\delta \right) - 4.322$ & $88$ & $8.2 \times 10^{2\, \dag \dag \dag}$ \\ \hline
Floating-Point$^{b}$ \cite{FloatingPoint} & $8\, \mathrm{log}_2 \left(1/\delta\right)+1.14 \, \mathrm{log}_2 \left(10^{\gamma}\right)$ & $>2.9 \times 10^2$ & $>2.9 \times 10^3$ \\ \hline
Ladder States$^{c}$ \cite{DistillNonstabState} &  -  & $1.1\times 10^{2}$ & $4.3 \times 10^3 $ \\ \hline
CISC$^{c}$ \cite{CISCLessMagic} & - & $7$ & $34$ \\ \hline
RS$^{a}$ \cite{OptimalCliffordTDecompOfZ} & $3 \, \mathrm{log}_{2} \left(1/\delta \right) + \mathrm{log}_{2} \left(\mathrm{log}_{10} \left(1/\delta \right) \right)$ & $1.1 \times 10^{2}$ & $1.1 \times 10^3$ \\ \hline
RUS$^{b}$ \cite{RUSSingleQubit, EffSynthRUS} & $1.15 \, \mathrm{log}_2 \left(1/\delta \right)$ & $42$ & $4.2 \times 10^{2}$ \\ \hline
DCP$^{c}$ \cite{OneStepComplexDistillation} & - & $7$ &$34$ \\ \hline
PQF$^{b}$ \cite{PQFterminates} & $\mathrm{log}_2 \left(1/\delta \right) + \, \mathrm{log}_{10} \left( \mathrm{log}_{10} \left(1/\delta \right) \right)$ & $38$ & $3.7 \times 10^{2}$ \\ \hline
\end{tabular}
\caption{\label{tbl1} Resource states required for non-Clifford gates for the various methods discussed in the review.  Techniques are presented approximately chronologically.  Two assessments of the number of non-Clifford states are presented: a target gate precision $\left(\delta\right)$ of $10^{-2}$ in the left column and $\delta = 10^{-20}$ on the right except for the Fowler $\left( \dag \right)$, PARs $\left( \dag \dag \right)$, and KMM $\left( \dag \dag \dag \right)$ which are evaluated at a precision of $10^{-6}$, $10^{-5}$, and $10^{-15}$, respectively.  The lower bounds on the floating-point method are given as the $\hat{T}$ gate assessment target is angle dependent ($\theta = a \times 10^{-\gamma}$).  The superscripts denote the category of the technique as presented in the text: $\left( a \right)$ Clifford and $\hat{T}$ decompositions, $\left( b \right)$ non-determinisitc methods, $\left( c \right)$ complex distillation.  The estimates above assume the availability of sufficiently precise magic states. This assumption increases the advantage of complex distillation techniques. The Ladder State value relative to CISC and DCP is inflated as the number includes the error-free distillation cost of the higher-order states. The advantage of DCP over CISC is in the distillation procedures for noisy magic states (see text).  The scaling factors $s_1$ and $s_2$ depend on the specific algorithm but are $\gtrsim$ 1.}
\end{table}

\section*{\sffamily \Large FUTURE DIRECTIONS}

Without functioning devices, adequate estimation of resource costs of quantum simulation remains difficult.  When one considers the wide use of heuristics and approximation in quantum chemistry by classical computation, it seems likely that a similar explosion of methods will occur when a real device is present. 

The advantage of a quantum chemical calculation on a quantum computer is the low cost of expanding configuration space.  The disadvantage is that, for the canonical algorithm, the energies are calculated by solving the dynamics problem. This suggests that quantum computers may have a larger advantage for calculating dynamic properties \cite{PolynomTimeQuantAlgoChemDyn} and also that new algorithmic ideas, such as variational solvers \cite{VarEigenSolver}, may lead to better ways to retrieve energies.  

We have outlined resource efficient methods of approximating arbitrary qubit rotations that will be required in computing the elements of the Trotter decomposition of the time evolution of a molecular Hamiltonian.  An open question is how to decompose, within a given precision, the continuous time evolution of the Hamiltonian into a set of Trotter steps in a manner which is efficient (logarithmic or polylogarithmic) in resources and time during the simulation.  Algorithms have been developed which have displayed such decompositions and exploit trade-offs between the number of qubits and the number of parallel computational steps \cite{BerrySparseH,GateEffQuantumQuery,ExpImrovTimeEvol,ExpImprovSparseH}. Connecting the Trotter compilation to rotation composition and distillation could yield further reductions in the total number of gates, a notion partially investigated by Hastings et al. \cite{ImproveQuantAlgoQuantSim}.  Furthermore, an understanding of the required precision of the rotations implemented in the quantum chemistry algorithms to achieve meaningful results is critical for implementation into physical devices.

\subsection*{\sffamily \large ACKNOWLEDGMENTS}


We thank Adam Meier, Rob Parrish, and Mauricio Guti\'{e}rrez-Arguedas for helpful comments on the manuscript. We also thank MGA for helping to prepare figure \ref{fig:circuitidentities}. This work was supported by NSF PHY-1415461.  


\bibliography{bibtexrefs.bib}   

\section*{\sffamily \Large APPENDIX}

In this section we will show how to map the dynamics of the molecular Hamiltonian (Equation \ref{eq:MolHamilt}), more specifically an element of the Trotter decomposition, onto a set of Clifford operations and arbitrary qubit rotations.  Given a Trotter step $\exp\left(-\frac{it}{\hbar}\hat{\mathcal{H}}_k/m\right)$, we can utilize the following identity:  Let $\hat{g}$ be the Clifford operator such that $\hat{g} \hat{Z}_j \hat{g}^\dagger$=$\hat{P}$, where $\hat{P}$ is a Hermitian Pauli operator ($\hat{Z}_i\hat{Z}_j\hat{X}_k\hat{X}_l$ for example), then the unitary $\hat{R}_P(\theta)=\exp(-i \theta \hat{P} /2)=\hat{g}\exp(-i \theta \hat{Z}_i /2 ) \hat{g}^\dagger=\hat{g}\hat{R}_{Z_j}(\theta)\hat{g}^\dagger$, where $\hat{R}_{Z_j}(\theta)$ is the rotation of qubit $j$ about the $z$-axis by an angle $\theta$.  The key result is that the parameters of the Hamiltonian are encoded into the simulation by single-qubit rotations. 

We no show how to find the Clifford operator $\hat{g}$ that transforms a rotation into a piece of the molecular Hamiltonian.  Consider the term $h_{jk}(\hat{c}^\dagger_j\hat{c}_k+\hat{c}^\dagger_k\hat{c}_j)$ in the molecular Hamiltonian where $h_{jk}$ is real. We wish to implement the unitary operator $\exp \left(-i (h_{jk}t/\hbar)\left[\hat{c}^\dagger_j\hat{c}_k+\hat{c}^\dagger_k\hat{c}_j\right]\right)$  for some time $t$.  Utilizing the Jordan-Wigner transformation, we have \cite{SimElecStruct}:
\begin{equation*}
\hat{c}^\dagger_j = (\hat{X}_j-i\hat{Y}_j)\prod_{l=j+1}^N\hat{Z}_l.
\end{equation*}
The long chain of $\hat{Z}$ operators guarantees the correct anticommutation behavior.
Without loss of generality, assume $k>j$ then 
\begin{equation*}
\hat{c}^\dagger_j\hat{c}_k=(\hat{X}_j-i\hat{Y}_j)\left(\prod_{l=j+1}^{k-1}\hat{Z}_l\right)(\hat{X}_k+i\hat{Y}_k) 
\end{equation*}
and we see that 
\begin{equation*}
 \hat{c}^\dagger_j\hat{c}_k+\hat{c}^\dagger_k\hat{c}_j =\hat{X}_j\left(\prod_{l=j+1}^{k-1}\hat{Z}_l\right)\hat{X}_k+\hat{Y}_j\left(\prod_{l=j+1}^{k-1}\hat{Z}_l\right)\hat{Y}_k.
\end{equation*}
is the sum of two Pauli operators that commute. 

The next step is to build the Clifford operators to transform a single-qubit $\hat{Z}$ operator into these $\left(k-j+1 \right)$-body spin operators. We use three identities: $\hat{\mathrm{CNOT}}(l,m)\hat{Z}_m\hat{\mathrm{CNOT}}(l,m)=\hat{Z}_l\hat{Z_m}$ allows us to grow a chain of $\hat{Z}$ operators, $\hat{H}\hat{Z}\hat{H}=\hat{X}$ transforms $\hat{Z}$'s to $\hat{X}$'s, and $\hat{S}\hat{H}\hat{Z}\hat{H}\hat{S}^\dagger=\hat{Y}$ transforms $\hat{Z}$'s to $\hat{Y}$'s.  

We apply these identities to the problem of finding a Clifford operator $\hat{g}_{XX}$ that satisfies the equation $\hat{g}_{XX}\hat{Z}_k\hat{g}_{XX}^\dagger=\hat{X}_j\left(\prod_{l=j+1}^{k-1}\hat{Z}_l\right)\hat{X}_k$. The solution is to grow a chain of $\hat{Z}$'s from $k$ to $j$, $\left(\prod_{l=j}^{k-1}\hat{\mathrm{CNOT}}(l,l+1)\right)^\dagger \hat{Z}_k\left(\prod_{l=j}^{k-1}\hat{\mathrm{CNOT}}(l,l+1)\right)=\prod_{l=j}^k\hat{Z}_l$, and then apply two $\hat{H}$'s; $\hat{g}_{XX}=\hat{H}_j\hat{H}_k\left(\prod_{l=j}^{k-1}\hat{\mathrm{CNOT}}(l,l+1)\right)^\dagger$. A similar analysis for the other Pauli operator reveals $\hat{g}_{YY}=\hat{S}_j\hat{S}_k\hat{g}_{XX}$.

Combining all of the pieces we have
\begin{equation*}
\exp \left(-i (h_{jk}t/\hbar)\left[\hat{c}^\dagger_j\hat{c}_k+\hat{c}^\dagger_k\hat{c}_j\right]\right)=\hat{g}_{XX}\hat{R}_{Z_k}(2h_{jk}t/\hbar)\hat{g}^\dagger_{XX}\hat{g}_{YY}\hat{R}_{Z_k}(2h_{jk}t/\hbar)\hat{g}^\dagger_{YY}
\end{equation*}
which is a unitary operator made from Clifford operations and a rotation about the $z$ axis set by the molecular Hamiltonian and the time of simulation. Although the details of the transformation differ, this gate set is also all that is required for the Bravyi-Kitaev mapping \cite{BKtransformQuantChem}. 

\end{document}